\documentclass[prd,aps,nofootinbib,twocolumn,amsmath,amssymb,preprintnumbers]{revtex4-1}
\usepackage{natbib}
\usepackage{amsmath}
\usepackage{amsfonts}
\usepackage{amssymb}
\usepackage[utf8]{inputenc}
\usepackage{bbm}
\usepackage{epsfig}
\usepackage{psfrag}
\usepackage{subfigure}
\usepackage{graphics}
\usepackage{graphicx}

\newcommand{\dd}{\mathrm{d}}

\renewcommand{\vec}[1]{\ensuremath{\mathbf{#1}}}

\begin{document}

%
%

\title{Peculiar Velocity Anomaly from Forces Beyond Gravity?}
\author{Youness Ayaita, Maik Weber, Christof Wetterich}
\affiliation{Institut f\"ur Theoretische Physik,
Universit\"at Heidelberg\\
Philosophenweg 16, D-69120 Heidelberg, Germany}

%
%

\begin{abstract}
	We address recently reported anomalously large bulk flows on
	scales of $100 h^{-1}$Mpc and beyond. These coherent motions of
	galaxies challenge the standard $\Lambda$CDM concordance model as
	well as a large class of competitive models of dark energy and
	modified gravity. If confirmed, they may support alternative
	models that include extra couplings inducing enhanced peculiar
	velocities on large scales. A complementary probe of the evolution
	of large--scale perturbations is the integrated Sachs--Wolfe
	effect; we explore the connection between this observable and the
	bulk flow. For illustration, we consider a coupling between dark
	energy and dark matter as well as a specific cosmological model,
	growing neutrino quintessence.
\end{abstract}

\maketitle

%
%

\section{Introduction}
\label{introduction}

Recent observations of large--scale galaxy motions constitute one of
the main challenges for the cosmological standard model
\citep{Watkins08, Perivolaropoulos08, Afshordi08}. In a Gaussian
window of diameter $100 h^{-1}$Mpc, \citet{Feldman09} find a bulk
motion of $416\pm 78$~km/s in conflict with the expected variance of
$\approx 200$~km/s at the $2\sigma$ level. Other analyses of the
peculiar velocity field do not draw a coherent picture \citep{Song10a,
Macaulay10, Colin10, Song10b, Osborne10, Dai11, Kashlinsky08,
Kashlinsky09a, Kashlinsky09b, Kashlinsky10, AtrioBarandela10}.  While
some observations do not detect anomalous flows \citep{Song10b,
Osborne10, Dai11}, other results confirm the presence of unexpectedly
large bulk motions \citep{Kashlinsky08, Kashlinsky09a, Kashlinsky09b,
Kashlinsky10, AtrioBarandela10}. \citet{Kashlinsky08, Kashlinsky09a}
investigate scales of $\approx 300 h^{-1}$Mpc, where the expectation
is even lower, obtaining the drastic result of $600$--$1000$~km/s.
Despite of the large uncertainties still present today, such values
have the potential of forming a highly significant anomaly for the
$\Lambda$CDM model in the future.

Yet, the standard $\Lambda$CDM model has passed a series of stringent
tests. It assumes a spatially flat universe essentially made up from
dark energy in the form of a cosmological constant $\Lambda$, cold
dark matter (CDM), baryons, and radiation. According to the standard
picture, matter perturbations from a nearly scale--invariant
primordial spectrum grew solely due to Einstein gravity. Since the
peculiar velocity field is intimately connected to the growth of
structure, a modified growth history will typically affect the
expected peculiar velocities. This can occur, e.\,g., in models of
modified gravity \citep{Amendola99a, Carroll04, Starobinsky07,
Carloni07, Tsujikawa07, Gong08}, brane--world models \citep{Dvali00,
Cardoso07, Khoury09}, and models including extra
couplings between the dark components \citep{Wetterich95, Amendola99b,
Brookfield05, Amendola07, Wetterich07, Ichiki08, Mota08,
CalderaCabral09, Gavela09}. Some authors have already addressed the
anomalous bulk flow in the context of specific models
\citep{MersiniHoughton08, Afshordi08, Jimenez08}.

The evolution of large--scale perturbations also leaves an imprint on
the temperature anisotropies of the cosmic microwave background (CMB)
by virtue of the integrated Sachs--Wolfe effect (ISW). This imprint is
seen in the cross--correlation between temperature anisotropies and
large--scale structure \citep{Ho08, Giannantonio08}. These
observations are sensitive to similar scales as the observations of
bulk flows \citep{Watkins08} and are thus an important complementary
probe.

In the following, we first point out, with the help of perturbation
theory, why the peculiar velocity anomaly not only challenges
$\Lambda$CDM, but also a large class of competitive models
(Sec.~\ref{linear_theory}). We then turn to the close relationship
between the bulk flow and ISW observables (Sec.~\ref{bulk_isw}).
Based on the considerations of Secs.~\ref{linear_theory} and
\ref{bulk_isw}, we investigate two scenarios illustrating potential
ways of addressing the anomaly in Sec.~\ref{examples}. We conclude
in Sec.~\ref{conclusion}.

%
%

\section{Bulk Flows in Linear Perturbation Theory}
\label{linear_theory}

We begin with the description of peculiar velocities in perturbation
theory from which we then infer that large bulk flows are unexpected
in $\Lambda$CDM and in most alternative models. Inhomogeneities in
the metric induce deviations from the uniform Hubble flow. They are
accounted for by the peculiar velocity field $\vec v(\vec x)$. The
bulk flow $\vec u$ is the average peculiar velocity in some volume
defined by a window function $W$,
\begin{equation}
    \vec u(\vec x) = \int_{}^{} \dd^3 y \, \vec v(\vec y)\, W(\vec
    x-\vec y).
    \label{bulk_def}
\end{equation}
Every cosmological model predicts the mean square $\left< u^2 \right>$
for a window of given size and shape, which can be compared with
observation. Throughout, we stick to statistical homogeneity and
isotropy, which for the Fourier transformed velocity field $\vec
v_{\vec k}$ implies
\begin{equation}
    \left< \vec v_{\vec k}^* \vec v_{\vec k'} \right>
    = (2\pi)^3 P_v(k)\,\delta^3(\vec k - \vec k'),
    \label{Pv_def}
\end{equation}
where we have introduced the peculiar velocity power spectrum
$P_v(k)$. It enables us to write
\begin{equation}
    \left< u^2 \right>
    = \frac{1}{2\pi^2} \int_{0}^{\infty} \dd k\, k^2 P_v(k)
	|\tilde W(k)|^2
    \label{u_exact}
\end{equation}
for a Fourier transformed spherically symmetric window $\tilde W(k)$.

It is instructive to relate the peculiar velocity power spectrum
$P_v(k)$ to the matter density power spectrum $P_\delta(k)$. This is
achieved with the aid of the continuity equation. In the Newtonian
limit, it reads
\begin{equation}
	\dot \delta_k = - k\, v_k,
	\label{continuity}
\end{equation}
where a dot denotes the derivative with respect to conformal time, and
$v_k$ is the scalar velocity perturbation. We define the average
growth factor $f_k = {\dd \log \delta_k}/{\dd \log a}$ (being
independent of $k$ in the standard picture), such that
Eq.~(\ref{continuity}) reads
\begin{equation}
	\mathcal H\, f_k\, \delta_k = - k\,  v_k,
    \label{continuity_f}
\end{equation}
with the conformal Hubble parameter $\mathcal H = \dot a / a$.
Introducing the spectra yields
\begin{equation}
	P_v(k) = \frac{f_k^2\, {\mathcal H}^2}{k^2} P_\delta(k).
    \label{Pv_Pdelta}
\end{equation}
Equation~(\ref{Pv_Pdelta}) tells us that larger bulk flows demand
higher values of $f_k$ or $P_\delta(k)$. This, however, poses a
serious obstacle for most cosmological models that reproduce the
standard expansion history. On the one hand, once a model is chosen,
the density power spectrum $P_\delta(k)$ is constrained by various
observations (like the CMB and galaxy surveys). Therefore, most
cosmological models do not allow for drastic deviations from the
$\Lambda$CDM power spectrum. On the other hand, for a large class of
dark energy and modified gravity models, $f_k$ can be parameterized by
$f_k = \Omega_m^\gamma$ with $\gamma$ constant in time
\citep{Linder05}. \citet{Linder07} showed that for models of
uncoupled dark energy, $\gamma$ only slightly depends on the equation
of state $w$, and that even when considering models of modified
gravity, $\gamma$ typically varies at most $\approx 20\%$, not enough
to predict the observed bulk flows. Consequently, what at first was
found as a challenge for the standard model $\Lambda$CDM, is in fact a
problem for its most popular competitors as well.

This should come as no surprise since the direct influence of
uncoupled dark energy, whether it be dynamical or a cosmological
constant, is restricted to the evolution of the background. Similar
expansion histories thus imply similar growth histories. This
correspondence is absent in models with extra couplings. In fact,
these models have the potential of generating large bulk flows. We
will illustrate this with two scenarios in Sec.~\ref{examples}.

An alternative approach would be to alter the primordial spectrum of
perturbations in order to obtain the observed bulk flows without
abandoning the standard dynamics. Regarding the lack of large--scale
power in the CMB maps \citep{Hajian07, Copi08, Ayaita09}, however, it
would seem more natural not to assume enhanced primordial power on the
largest scales. We thus concentrate on modified dynamics.

Even in case a model dynamically accounts for large bulk flows, it
also has to satisfy the constraints from ISW observations. We will
now turn to this complication.

%
%

\section{Relation Between Bulk Flow and ISW Observables}
\label{bulk_isw}

A very related observable to the large--scale bulk flow is given by
the ISW \citep{Watkins08} as measured in the cross--correlation
between temperature anisotropies and matter perturbations \citep{Ho08,
Giannantonio08}. The ISW temperature anisotropy $\Delta T^\text{ISW}$
is the consequence of CMB photons traversing time--varying
gravitational potentials along their path,
\begin{equation}
	\frac{\Delta T^\text{ISW}}{T} = 2\,
	\int_{}^{} \dot \Phi \, \dd \tau.
	\label{ISW}
\end{equation}

\citet{Watkins08} already referred to the interesting fact that the
ISW observations ---~sensitive to similar scales as the bulk flow
measurement~--- are not in good agreement with the $\Lambda$CDM
best--fit model either. This raises the question whether the two
independent results are related to each other. We will see, however,
that in the standard picture, both results point to opposite
directions.

The ISW effect can be detected by measurements of the
cross--correlation between temperature and matter density
fluctuations. The observable thus depends on the product $\dot \Phi_k
\delta_k$; since $\delta_k$ is constrained by large--scale structure
observations, the interesting contribution comes from $\dot \Phi_k$.
This motivates the definition of an ISW amplitude, normalized to the
fiducial $\Lambda$CDM model,
\begin{equation}
	A_k(z) = \frac{\dot \Phi_k(z)}{\dot \Phi^\text{fid}_k(z)}.
	\label{A_ISW}
\end{equation}
The ISW measurement of \citet{Ho08} provides an observational result,
which is effectively averaged over a range of scales and redshifts,
$\bar A^\text{obs}=2.23 \pm 0.60$, about $2\sigma$ above the
$\Lambda$CDM expectation (for which trivially $\bar A = 1$). The
observation takes most of its sensitivity from scales $k\approx 0.01$
to $0.03\,h/\text{Mpc}$ at redshift $z_\text{ISW}\approx 0.5$
\citep{Ho08, Afshordi08}. Hence, for a cosmological model, we expect
good agreement with the observation if $A_k(z_\text{ISW})\approx \bar
A^\text{obs}$ at these scales and redshift.

The derivative $\dot \Phi_k$ of the gravitational potential is also
related to the velocity perturbation $v_k$. Combining the continuity
equation~(\ref{continuity}) with the Poisson equation $k^2 \Phi_k = -
3/2\ \mathcal H^2 \Omega_m \delta_k$, we obtain
\begin{equation}
	v_k = \frac{2 k}{3 \mathcal H^2 \Omega_m} \left( \mathcal H \Phi_k +
	\dot \Phi_k \right).
	\label{bulk_phi}
\end{equation}
Due to the accelerated expansion, the large--scale gravitational
potential decays if there is no coupling beyond Einstein gravity. The
two terms $\mathcal H \Phi_k$ and $\dot \Phi_k$ thus have opposite
signs and partially cancel. At the present cosmic time, the
$\Lambda$CDM best--fit model predicts $\mathcal H \Phi_k / \dot \Phi_k
\approx -2$ at scales typical for the bulk flow observation. Hence, we
can infer two possibilities to generate larger peculiar velocities
from Eq.~(\ref{bulk_phi}). The first would require deeper potentials
$\Phi_k$, the second a change in $\dot \Phi_k$ such that the
cancellation of the two contributions is reduced. Since the potential
$\Phi_k$ is constrained by large--scale structure observations of the
recent universe, we expect the contribution $\propto \dot \Phi_k$ to
be decisive for a potential explanation of the large bulk flow. We
thus have a look at the contribution from the time evolution of the
gravitational potential,
\begin{equation}
	v^\text{ev}_k = \frac{2k}{3\mathcal H^2 \Omega_m}\,
	\dot \Phi_k.
	\label{ev}
\end{equation}
This contribution leads to larger bulk flows if it is of smaller
magnitude than in the $\Lambda$CDM case, i.\,e.\/ for a slower decay
(or an increase) of the gravitational potential. In fact, for a
constant gravitational potential implying $v_k^\text{ev} = 0$, we
would, from Eqs.~(\ref{bulk_phi}) and (\ref{u_exact}), already get a
bulk flow variance $U = \sqrt{\left< u^2 \right>} \approx
440~\text{km/s}$.

A decay of the large--scale gravitational potential and hence a
non--vanishing $\dot \Phi_k$, however, is not only a model prediction
but actually observed in ISW measurements. More formally, the
connection between the ISW signal amplitude $A_k$ and the velocity
perturbation is obtained by inserting Eq.~(\ref{A_ISW}) into
Eq.~(\ref{ev}),
\begin{align}
	v_k^\text{ev}(z=0) &= 
	\frac{2
	k\,\dot\Phi^\text{fid}_k(z=0)}{3
	H_0^2 \Omega_m^0}
	\, q_k \, A_k(z_\text{ISW}) \nonumber\\
	& \equiv
	\alpha_k \, q_k \, A_k(z_\text{ISW}),
	\label{ev_with_A}
\end{align}
where the factor $q_k$ quantifies the change of $\dot \Phi_k$ between
$z = z_\text{ISW} \approx 0.5$ and $z = 0$. More formally, it is given
by the quotient $\dot \Phi_k(z = 0) / \dot \Phi_k(z_\text{ISW})$
divided by the corresponding value for the fiducial $\Lambda$CDM model
(for which $q_k = 1$). It accounts for the fact that the ISW
observation, most sensitive at $z_\text{ISW}$, probes slightly earlier
times than the bulk flow observation at $z = 0$. The quantities
entering the factor $\alpha_k$ are tightly constrained. At scales of
about $100~\text{Mpc}$, i.\,e. $k \approx 0.01/\text{Mpc}$, it amounts
to $\alpha_k \approx 220~\text{km/s}$.

The observed ISW signal amplitude ---~itself about $2\sigma$ away from
the $\Lambda$CDM case~---, $\bar A^\text{obs} \approx 2$, suggests a
decay of the gravitational potential about twice as fast as in the
fiducial $\Lambda$CDM case~\citep{Ho08}. If, for illustration, we
assumed this value on all large scales at the present cosmic time, the
larger cancellation $v_k^\text{ev}$ would reduce the bulk flow
variance $U = \sqrt{\left< u^2 \right>}$ to substantially less than
$100~\text{km/s}$. This illustrates the tension between the ISW and
bulk flow observations. If the gravitational potential indeed decays
that fast, the peculiar velocity anomaly looks even more severe. Since
the two observations, ISW effect and bulk flow, suppose opposite
behaviors of the large--scale gravitational potential, it is
impossible to alleviate both disagreements within the standard
framework. This tension, if confirmed, may motivate more complex
cosmological dynamics.

%
%

\section{Examples of Extra Couplings}
\label{examples}

In Sec.~\ref{linear_theory}, we concluded that models with extra
couplings are candidates capable of generating large bulk flows.
Since the most prominent large--scale structure observations are in
good agreement with the $\Lambda$CDM model, we expect the large bulk
flows to be a recent phenomenon that has not yet had a significant
impact on the matter density fluctuations. It is natural to suspect
a link to the recent onset of dark energy domination. For this reason,
we consider models of coupled dark energy. In a first scenario, we
assume a coupling between dark energy and cold dark matter; in a
second scenario, dark energy is coupled to another component,
neutrinos in this case, and the effect on cold dark matter is only
indirect.

For both scenarios, we describe dark energy with a dynamical scalar
field $\varphi$, the cosmon \citep{Wetterich88, Ratra88}. Its time
evolution is given by the Klein--Gordon equation in presence of a
cosmon potential. An often--used example is the exponential potential
(cf., e.\,g., Ref.~\citep{Wetterich08}),
\begin{equation}
	V(\varphi) \propto \exp(-\alpha \varphi),
	\label{}
\end{equation}
where $\alpha$ is a dimensionless parameter of the model for which
constraints on early dark energy suggest $\alpha \gtrsim 10$
\citep{Doran07}. Here and in the following, we use units where $8\pi G
= 1$.

A coupling between dark energy and a species $A$ (dark matter in the
first and neutrinos in the second scenario) means the exchange of
energy and momentum between both fluids. The individual
energy--momentum tensors are no longer conserved, but only their sum
is,
\begin{equation}
	\nabla_\nu T_A^{\mu\nu} = Q^\mu, \
	\nabla_\nu T_\varphi^{\mu\nu} = - Q^\mu.
	\label{exchange}
\end{equation}
No symmetry is known that would enforce $Q^\mu = 0$ whence, in
general, we have to a expect a non--vanishing coupling. The specific
form of the couplings investigated in this work was proposed by Refs.
\citep{Wetterich95, Amendola99b},
\begin{equation}
	Q^\mu = - \beta \, {\left( T_A \right)^\nu}_\nu\, \nabla^\mu \varphi.
	\label{Q}
\end{equation}
The coupling parameter $\beta$ may be constant in time or depend on
the cosmon, $\beta = \beta(\varphi)$. The modified conservation
equations~(\ref{exchange}) imply additional terms in the evolution of
the background densities,
\begin{align}
	\dot \rho_\varphi
	&= - 3 \mathcal H (1 + w_\varphi)\rho_\varphi
	+ \beta \, \dot\varphi (1 - 3 w_A) \rho_A,
	\label{bg1}
	\\
	\dot \rho_A
	&= - 3 \mathcal H (1 + w_A) \rho_A
	- \beta \, \dot\varphi (1 - 3 w_A) \rho_A.
	\label{bg2}
\end{align}
In the perturbation equations, the coupling mediates an extra force
between particles of species $A$ leading to stronger structure
formation allowing for larger peculiar velocities.

%
%

\subsection{Dark Energy Coupled to Cold Dark Matter}
\label{dedm}

As a first scenario, we assume a non--vanishing coupling $\beta$
between dark energy and cold dark matter, while the couplings of dark
energy to other matter species are assumed to be negligible. The
background evolution is given by Eqs.~(\ref{bg1}) and (\ref{bg2})
inserting the equation of state $w_A \equiv w_c = 0$. With an
appropriate choice of the cosmon potential, it leads to an expansion
history similar to $\Lambda$CDM.

In the presence of a non--vanishing coupling, the evolution of dark
matter perturbations in the Newtonian limit \citep{Amendola99b} is
given by
\begin{align}
	\ddot \delta_{c,k} &+ (\mathcal H - \beta \dot \varphi) \,
	\dot \delta_{c,k} \nonumber \\
	&- \frac{3}{2} \mathcal H^2
	\left[ 
	\left( 1 + 2 \beta^2 \right) \Omega_c \delta_{c,k}
	+ \Omega_b \delta_{b,k}
	\right] = 0.
	\label{dedmpert}
\end{align}
The major effect of the coupling visible in this equation is the
modified force term, which for dark matter is enhanced by a factor of
$1+2\beta^2$ compared to the uncoupled case. The coupling~(\ref{Q})
thus mediates an additional attractive force between dark matter
particles that can be modeled by an effectively enhanced Newton's
constant, $G_\text{eff} = (1+2\beta^2) G$. A direct consequence are
larger peculiar velocities in the dark matter fluid. Under the
assumption that tracers like galaxies follow the dark matter
distribution, these flows could be measured.

The second effect in Eq.~(\ref{dedmpert}) is the modification of the
damping term $\propto \mathcal H - \beta \dot \varphi$. Since an
accelerated expansion is only obtained for an effectively stopped
evolution of the cosmon, the term $\beta \dot \varphi$ is expected to
be negligible if $\beta$ is of order unity or smaller. The term is
already small during matter domination where the scaling solution
tells us that $\dot \varphi$ is of order $\mathcal H/\alpha \ll
\mathcal H$ \citep{Wetterich95}.

The simple idea, however, of generating large peculiar velocities
through a strong direct coupling between dark energy and dark matter
suffers from several drawbacks. As explained in Sec.~\ref{bulk_isw},
larger peculiar velocities require a slower decrease or even an
increase of the large--scale gravitational potential. We have argued
that this is in conflict with ISW observations. A second complication
comes from the effect of the coupling on the matter density power
spectrum for which various large--scale structure observations provide
tight constraints.

These objections tell us that the coupling between dark energy and
dark matter, in order to remain consistent with observational data
while generating large peculiar velocities, needs to have a more
complex time evolution, realized, e.\,g., by a varying parameter
$\beta = \beta(\varphi)$. According to the continuity
equation~(\ref{continuity}), larger peculiar velocities today only
require an enhanced {\it present} growth of matter perturbations. If
this growth has set in very recently, the density power spectrum may
be affected only mildly. This motivates to consider the possibility of
a coupling $\beta$ being negligible in the far past and becoming
effective only in recent times. A more sophisticated theory of coupled
dark energy in which an extra force naturally becomes effective only
at late times is growing neutrino quintessence investigated in
Sec.~\ref{growing_neutrinos}. For now, however, we simply employ the
ad--hoc assumption of a step--like behavior of $\beta$. In our
numerical illustration, we set the trigger to the onset of dark energy
domination at $a \approx 0.4$.

\begin{figure}[htb]
	\begin{center}
		\subfigure
		{
			\psfrag{xlabel}[B][c][0.95][0]{Late--time coupling $\beta$}
			\psfrag{ylabel}[B][c][0.95][0]{Bulk flow variance $U$ [km/s]}
			\includegraphics[width=.4\textwidth]{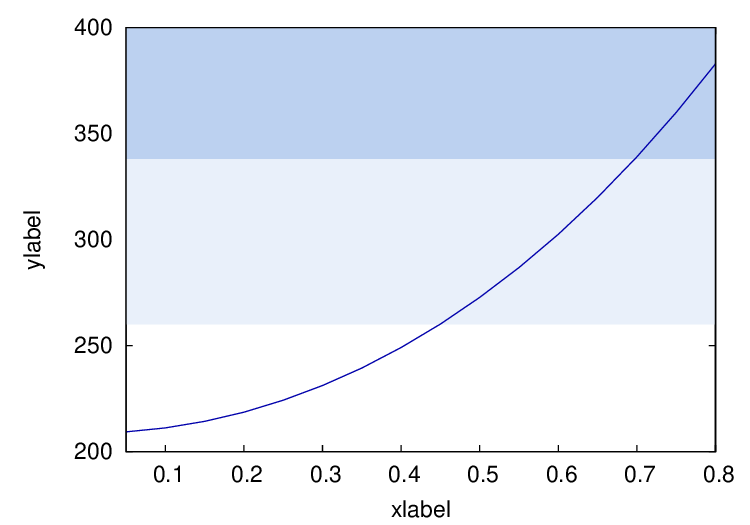}
			\label{dedm_bulk}
		}
		\subfigure
		{
			\psfrag{xlabel}[B][c][0.95][0]{Late--time coupling $\beta$}
			\psfrag{ylabel}[B][c][0.95][0]{ISW signal amplitude
			$A$}
			\includegraphics[width=.4\textwidth]{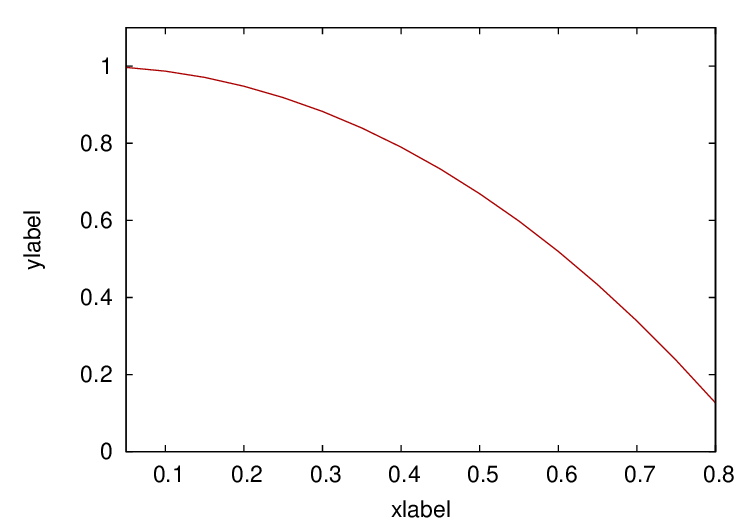}
			\label{dedm_isw}
		}
	\end{center}
	\caption{The effect of the coupling $\beta$ between dark
	energy and dark matter on the bulk flow and ISW expectations. The
	shaded regions mark the observational $1$~and $2\,\sigma$
	intervals.}
	\label{dedm_plot}
\end{figure}

We numerically integrate the perturbation equation~(\ref{dedmpert})
neglecting the small contribution of $\dot \varphi$ to the damping
term. For an illustration of the basic effects, it is sufficient to
take the Hubble parameter $\mathcal H$ and the initial values for
$\delta_{c,k}$ and $\dot \delta_{c,k}$ from the $\Lambda$CDM best--fit
model. In this manner, the evolution reduces to the $\Lambda$CDM case
for $\beta = 0$. Using Eqs.~(\ref{u_exact}) and (\ref{continuity}), we
can calculate the new bulk flow variance $U = \sqrt{\langle u^2
\rangle}$ in a Gaussian window of diameter $100 h^{-1}\text{Mpc}$
corresponding to the observation of \citet{Feldman09}. We further
estimate the ISW signal amplitude $A_k(z)$ according to
Eq.~(\ref{A_ISW}) at the scale and redshift named there. The results
for varying $\beta$ are shown in Fig.~\ref{dedm_plot}.

\begin{figure}[htb]
	\begin{center}
		\psfrag{xlabel}[B][c][0.95][0]{Scale factor $a$}
		\psfrag{ylabel}[B][c][0.95][0]{Gravitational potential
		$|\Phi_k|$}
		\includegraphics[width=.4\textwidth]{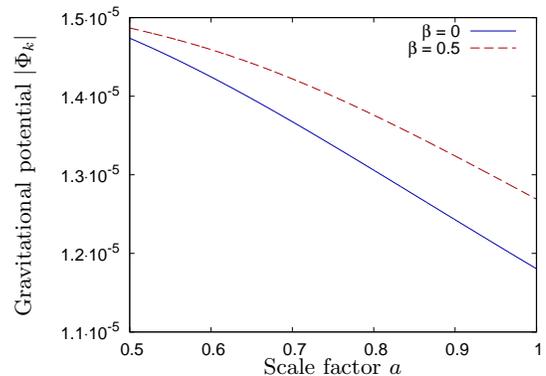}
		\caption{The time evolution of the gravitational potential (at
		the scale $k \approx 0.02 h/\text{Mpc}$ relevant for the ISW
		observation) for the uncoupled case, $\beta = 0$, and a
		coupling $\beta = 0.5$.}
		\label{dedm_phi}
	\end{center}
\end{figure}

The upper plot, Fig.~\ref{dedm_bulk}, shows that the extra force is
capable of generating large bulk flows for $\beta \lesssim 1$. In
contrast, the ISW amplitude, cf.~Fig.~\ref{dedm_isw}, decreases for
larger bulk flows illustrating the conflict explained in
Sec~\ref{bulk_isw}. This result is explained by the fact that the
enhanced bulk flow comes along with a slower decay of the
gravitational potential whose time evolution we show in
Fig.~\ref{dedm_phi}.

Although the model succeeds in generating large bulk flows, the
numerical results in accordance with the reasoning of
Sec.~\ref{bulk_isw} show a severe discrepancy with the observed ISW
amplitude. A way out could consist in a more complex time evolution of
$\beta$. If we had chosen an alternative onset of $\beta$ at $z <
z_\text{ISW} \approx 0.5$, the ISW signal amplitude would be less
affected. For appropriate values of $\beta$, large bulk flows could
still be reached as the factor $q_k$ would substantially differ from
unity in Eq.~(\ref{ev_with_A}).

%
%

\subsection{Growing Neutrino Quintessence}
\label{growing_neutrinos}

Growing neutrino quintessence is a cosmological model with an
interaction between the neutrinos and the cosmon $\varphi$
\citep{Amendola07, Wetterich07}. Its motivation is the coincidence
problem concerning the intriguing similarity of the fractional energy
densities in dark energy and matter just today. In realistic
scenarios, the cosmon--neutrino coupling $\beta$, defined in
Eq.~(\ref{Q}), is large compared to gravity such that the neutrinos
are important for the evolution of the cosmon although their
fractional energy density is small. The coupling becomes only
effective once the neutrinos become non--relativistic ($z_{nr} \approx
5\text{--}10$) and $w_\nu$ becomes substantially smaller than $1/3$ in
Eqs.~(\ref{bg1}) and (\ref{bg2}). After this trigger event, the
subsequent stop of the evolution of the cosmon leads to an onset of
dark energy domination at recent times, similar to $\Lambda$CDM. The
value of the cosmon potential at the time when its evolution stops is
associated to an effective cosmological constant, resulting in a
present fraction of dark energy expressed in terms of the present
average neutrino mass $m_\nu^0$:
\begin{equation}
	\Omega_\varphi^0 = - \frac{\beta}{\alpha}\frac{
	m_\nu^0}{16\,\text{eV}}.
	\label{de_fraction}
\end{equation}
It is in this way that the model addresses the coincidence problem.

The strong coupling $\beta$ has a drastic impact on the evolution of
perturbations in the neutrino fluid. In fact, large--scale stable
neutrino lumps form \citep{Brouzakis07, Mota08, Bernardini09,
Wintergerst09}. The gravitational potential of these lumps acts as an
extra source for the growth of dark matter perturbations. Due to this
indirect influence, dark matter perturbations can grow faster, leading
to, e.\,g., enhanced peculiar velocities. We discuss here the simplest
case which does not include any direct coupling between dark energy
and cold dark matter.

In growing neutrino quintessence, the neutrino mass varies depending
on the cosmon field, expressed by the coupling
\begin{equation}
	\beta = - \frac{\dd \ln m_\nu}{\dd \varphi},
	\label{beta_def}
\end{equation}
from which we obtain $m_\nu \propto \exp(-\beta \varphi)$ for constant
$\beta$ as chosen in this section; in principle, $\beta$ may be a
function of $\varphi$ \citep{Wetterich07}.

Once the neutrinos are non--relativistic, the perturbation
$\delta\varphi$ of the cosmon field mediates an attractive force
between the neutrinos of order $|\vec F| = |\beta\, \vec\nabla
\delta\varphi| = 2 \beta^2 |\vec F_\text{gravity}|$
\citep{Wintergerst09}. In realistic scenarios with an expansion
history similar to $\Lambda$CDM, we have $\beta^2 \gg 1$ so that the
extra force causes a rapid growth of neutrino perturbations becoming
non--linear at around $z_{nl} \approx 1\text{--}2$ and forming stable
lumps. Linear perturbation theory breaks down even on large scales.
The details of both the non--linear evolution and the final state are
not yet understood. At the current stage, quantitatively comparing the
model with observations and constraining its parameter space is hardly
possible.

Nonetheless, we will illustrate the model's potential of generating
enhanced bulk flows of matter. In general, we can think of two
alternative ways in which the model can account for the peculiar
velocity anomaly. First, if the cosmological gravitational potential
of neutrino lumps is sufficiently large, it may drive an enhanced
structure formation of matter. In the following, we will concentrate
on this possibility. A second and equally interesting possibility is
that the cosmological gravitational potential of neutrino lumps is too
small for having a strong influence on the growth of matter
perturbations in the cosmological average. Nonetheless, in the local neighborhood of a neutrino
lump, large peculiar velocities may occur. A Gaussian distribution of
fluctuations is no longer expected to be a good approximation.

Since a full analysis is not yet possible, we parameterize the main
characteristics (cf.\/ Refs.~\citep{Wintergerst09, Pettorino10}) of
the model as follows. We neglect the gravitational potential
$\Phi_{\nu,k}$ induced by perturbations in the neutrino fluid for the
evolution of cold dark matter until the non--linear evolution sets in
at $z_{nl}$. Since the neutrino lumps form very quickly thereafter
\citep{Mota08, Wintergerst09}, we assume that $\Phi_{\nu,k}$ is then
determined by a distribution of virialized lumps. In the limit of very
large scales, these lumps can approximately be described as
point--shaped, and if they are distributed randomly, the corresponding
gravitational potential is given by
\begin{equation}
	\Phi_{\nu,k} = \frac{\rho_\nu}{2\pi\, \sqrt{2 n}}\
	k^{-\frac{1}{2}}
	\label{phi_nu_physical}
\end{equation}
where $n$ denotes the comoving number density of neutrino lumps
\citep{Pettorino10}. Here, we have made several assumptions making
Eq.~(\ref{phi_nu_physical}) an upper bound. Namely, we have assumed
that all neutrinos are clustered in lumps and that the neutrino mass
inside the lumps grows, following the background evolution (although
backreaction effects may freeze $m_\nu$ inside the lumps
\citep{Pettorino10, Nunes11}). Moreover, on scales comparable to the typical
size of neutrino lumps or smaller, the potential will drop off faster
than $\propto k^{-1/2}$, depending on the lumps' density profile.
Although a more detailed treatment of the non--linear evolution might
predict a potential substantially smaller, we will stick with
Eq.~(\ref{phi_nu_physical}) as an upper bound.

The total gravitational potential essentially has two main
contributions, $\Phi_k = \Phi_{m,k} + \Phi_{\nu,k}$, neglecting the
gravitational potential of a clustered cosmon field. The evolution
equations for matter perturbations in the Newtonian limit, not
accounting for the non--Gaussian features of the non--linear
evolution, hence read
\begin{align}
	\dot \delta_{m,k} &= - k \, v_{m,k}, \label{gnqpert1} \\
	\dot v_{m,k} &= -\mathcal H v_{m,k} + k \, \left( \Phi_{m,k} +
	\Phi_{\nu,k} \right), \label{gnqpert2} \\
	k^2 \Phi_{m,k} &= - \frac{3}{2} \mathcal H^2 \Omega_m \delta_{m,k},
	\label{gnqpert3}
\end{align}
where $\Phi_{\nu,k}$ is taken from Eq.~(\ref{phi_nu_physical}) once
virialized neutrino lumps have formed. Since the continuity equation
and the Poisson equation (for the matter--induced potential
$\Phi_{m,k}$) are identical to the standard case, Eq.~(\ref{bulk_phi})
is unchanged. It allows to compute the peculiar velocity perturbation
$v_{m,k}$ in terms of $\dot \Phi_{m,k}$ and $\Phi_{m,k}$. An evolution
equation for $\Phi_{m,k}$ can be obtained by combining
Eqs.~(\ref{gnqpert1})--(\ref{gnqpert3}),
\begin{align}
	\ddot \Phi_{m,k} + & 3 \mathcal H \dot \Phi_{m,k}
	+ \left( \dot{\mathcal H} + 2 \mathcal H^2 - \frac{3}{2} \mathcal
	H^2 \Omega_m
	\right) \Phi_{m,k}
	= \nonumber \\
	& = \frac{3}{2} \mathcal H^2 \Omega_m \Phi_{\nu,k}.
	\label{phi_m_evolution}
\end{align}
We numerically integrate this equation starting at redshift $z_{nl}$
where virialized neutrino lumps have formed and $\Phi_{\nu,k}$ becomes
important; we assume $z_{nl} \approx 1.5$. We employ the model
parameters suggested by \citet{Pettorino10} for a constant
cosmon--neutrino coupling $\beta = -275$ corresponding to a
present--day neutrino mass of $m_\nu^0 = 0.48~\text{eV}$. The Hubble
parameter $\mathcal H$ as well as the initial values for $\Phi_{m,k}$
and $\dot \Phi_{m,k}$ are taken from the $\Lambda$CDM best--fit model.
In this manner, the evolution reduces to the $\Lambda$CDM case for
$\Phi_{\nu,k} = 0$; since the expansion history as well as the early
perturbation evolution in growing neutrino quintessence are close to
the $\Lambda$CDM case, the general behavior is unaffected.

\begin{figure}[htb]
	\begin{center}
		\psfrag{xlabel}[B][c][0.95][0]{Fraction $p$}
		\psfrag{ylabel}[B][c][0.95][0]{Velocity quotient
		$v_{m,k}/v_{m,k}^\text{$\Lambda$CDM}$}
		\includegraphics[width=.4\textwidth]{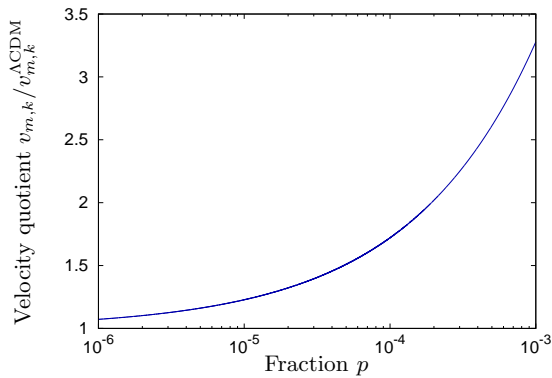}
	\end{center}
	\caption{Depending on the fraction $p$ of neutrinos bound in a
	single lump, the plot shows the amplification of the peculiar
	velocity on the scale $k = 0.01/\text{Mpc}$ characteristic for the
	bulk flow observation.}
	\label{gnq_plot}
\end{figure}

Since the bulk flow was observed at a scale of roughly
$100~\text{Mpc}$ \citep{Feldman09}, we consider a characteristic mode
$k \approx 0.01/\text{Mpc}$. For this mode, we compare the resulting
peculiar velocity $v_{m,k}$ from Eqs.~(\ref{bulk_phi}) and
(\ref{phi_m_evolution}) with the $\Lambda$CDM value
$v_{m,k}^\text{$\Lambda$CDM}$. The observed bulk flow exceeds the
$\Lambda$CDM expectation by a factor of about two,
$U^\text{obs}/U^\text{$\Lambda$CDM} \approx 2$; this suggests that
values $v_{m,k}/v_{m,k}^\text{$\Lambda$CDM}$ around $\approx 2$
indicate bulk flows in the right range. We vary the fraction $p$ of
the neutrinos in the Hubble volume concentrated in a single lump
(determining the number density of lumps $n \propto 1/p$). The
quotient $v_{m,k}/v_{m,k}^\text{$\Lambda$CDM}$ is shown in
Fig.~\ref{gnq_plot}.
We see a clear connection between the neutrino--induced gravitational
potential given by Eq.~(\ref{phi_nu_physical}), $\Phi_{\nu,k} \propto
\sqrt{p}$, and the amplification of the peculiar velocity of matter. Even the
values suggested by the observation of the bulk flow anomaly
\citep{Feldman09} can be reached.

Concerning the ISW, it would be desirable to also give an estimate
that could be compared with the observed signal amplitude.
Unfortunately, no robust estimation can be made at this stage. The
background density $\rho_\nu$ oscillates with time \citep{Wetterich07}
due to strong oscillations in the neutrino mass $m_\nu$. If these
oscillations visible in the background quantities also affect the
non--linear neutrino lumps, the neutrino--induced gravitational
potential will adopt this oscillatory behavior. Since its time
derivative $\dot \Phi_{\nu, k}$ enters the ISW estimation (assuming a
non--vanishing correlation between neutrino lumps and dark matter
structures), every prediction would be extremely sensitive to small
changes in the parameter values and to the details of the evolution.
Moreover, since our discussion of the ISW in Sec.~\ref{bulk_isw} is
based on linear perturbation theory, it remains open whether a full
non--linear treatment within growing neutrino quintessence would show
new and different phenomena relevant for the ISW observable.

On the other hand, there is the realistic possibility, already
mentioned above, that the neutrino--induced gravitational potential is
sub--dominant. In this case, the evolution of the large--scale
gravitational potential is dominated by matter and decays similarly to
the $\Lambda$CDM case. A neutrino lump in our cosmological vicinity
could nonetheless generate a large local bulk motion. The neutrino
fluid in growing neutrino quintessence is generically inhomogeneous on
large scales allowing for local features that are unlikely in the
standard scenario. The observed bulk flow could be such a phenomenon.

%
%

\section{Conclusion}
\label{conclusion}

In this work, we have investigated modifications to the standard
cosmological model that could be required if the peculiar velocity
anomaly persists. Extra couplings of the dark components are a natural
possibility, as we have illustrated with two scenarios. We have also
discussed the relationship between peculiar velocities and the ISW as
observed in the cross--correlation between temperature fluctuations of
the CMB and density fluctuations. Both are sensitive to the change of
the gravitational potential $\Phi_k$ with time in a similar range of
length scales $\propto k^{-1}$. While an enhanced bulk flow suggests a
slower decrease of $\Phi_k$ as compared to the fiducial $\Lambda$CDM
model, the observed ISW correlation requires an even faster decreasing
$\Phi_k$. In models with more complex cosmological dynamics, this
potential discrepancy can be alleviated by features in the time
evolution of $\Phi_k$ or if the large bulk flow is a local phenomenon.

As a first scenario, we have employed a coupling between dark energy
and dark matter, which succeeds in generating large peculiar
velocities but, in its simple form, conflicts with ISW observations.
Our results suggest that more elaborate cosmological models with a
dark coupling may be promising to resolve the peculiar velocity
anomaly if, at the same time, they respect further constraints.

We have also considered the case in which the additional growth is not
caused by an extra force directly acting on CDM particles, but rather
indirectly via an extra gravitational potential. This feature is
realized in growing neutrino quintessence, which includes a coupling
between dark energy and neutrinos. Under the influence of this extra
force, the neutrinos form large--scale structures in recent times,
generating an additional gravitational potential felt by the matter
perturbations. This scenario as well can be in accordance with the
peculiar velocity measurements.

If the significance of the peculiar velocity anomaly will increase
with future observations, it will be a strong sign for physics beyond
the $\Lambda$CDM concordance model. By our examples, we have seen that
modifications to $\Lambda$CDM becoming effective only in the recent
epoch may be sufficient to generate large peculiar velocities. This
might suggest that the new physics is connected with the recent
transition to dark energy domination.
\newline

%
%

We thank Bj{\"o}rn Malte Sch{\"a}fer for interesting and useful
discussions.

\bibliography{bulk}

\end{document}